\documentclass[a4paper]{article}

\usepackage[utf8]{inputenc}   
\usepackage[T1]{fontenc}      
\usepackage[english]{babel}  
\usepackage{amsmath}
\usepackage{parskip}
\usepackage{amssymb}
\usepackage{cite}
\usepackage{multicol}

\usepackage[squaren,Gray]{SIunits}
\usepackage{floatrow}

\usepackage[a4paper]{geometry}

\usepackage{lipsum}

\newcommand\blfootnote[1]{%
  \begingroup
  \renewcommand\thefootnote{}\footnote{#1}%
  \addtocounter{footnote}{-1}%
  \endgroup
}
                              
\usepackage{graphicx,wrapfig}
\usepackage[font=footnotesize,labelfont=bf]{caption}
\usepackage{float}                              

\title{Interfacial tension of reactive, liquid interfaces and its consequences}           
\author{Ana\"{i}s Giustiniani$^*$, Wiebke Drenckhan and Christophe Poulard$^*$ \blfootnote{ Corresponding authors e-mail addresses: anais.giustiniani@u-psud.fr (A. Giustiniani), christophe.poulard@u-psud.fr (C. Poulard).}}
\date{}

\begin{document}

\maketitle                   

\begin{abstract}
Dispersions of immiscible liquids, such as emulsions and polymer blends, are at the core of many industrial applications which makes the understanding of their properties (morphology, stability, etc.) of great interest. A wide range of these properties depend on interfacial phenomena, whose understanding is therefore of particular importance. The behaviour of interfacial tension in emulsions and polymer blends is well-understood -both theoretically and experimentally -in the case of non-reactive stabilization processes using pre-made surfactants. However, this description of the interfacial tension behaviour in reactive systems, where the stabilizing agents are created \textit{in-situ} (and which is more efficient as a stabilization route for many systems), does not yet find a consensus amongst the community. In this review, we compare the different theories which have been developed for non-reactive and for reactive systems, and we discuss their ability to capture the behaviour found experimentally. Finally, we address the consequences of the reactive stabilization process both on the global emulsions or polymer blend morphologies and at the interfacial scale.
\end{abstract}

\tableofcontents              

\section{Introduction}

Emulsions and polymer blends are dispersions of two immiscible liquids. In the case of emulsions, the liquids consist of low-molecular weight molecules, while in the case of a polymer blend, the two liquids are polymer melts which are often solidified after dispersion. Both, emulsions and polymer blends, find applications in number of industries in both liquid and solid state such as food, cosmetics, pharmaceutics, construction and plastics, thanks to their ability to combine the properties of both phases. In these biphasic systems, the morphology is a key parameter to understand and control the optical, rheological (in liquid/liquid state) and mechanical (in solid/liquid and solid/solid state) properties. However, the high interfacial tension between the liquids tends to separate the two phases, leading to coarse dispersions. This results in weak and anisotropic rheological and mechanical properties. Stabilization of the emulsions or polymer blends is then of capital interest, and requires the addition of a third constituent in the dispersion, called compatibilizer or surface active agent (shortened as surfactant). This component, either a short molecule, a polymer, a protein, or a particle, is dispersed in one of the phases (usually the continuous one) and diffuses towards the interface. The stabilization efficiency depends on the diffusion kinetics of the surfactant, on the visco-elastic properties of the surfactant monolayers created at the interface and on the interaction between the monolayers at the interface of two approaching dispersed drops.

There are two ways of adding surfactants to an emulsion or polymer blend. Either one can disperse the preformed surfactant in a separate step in one of the phases, or the surfactant can be created \textit{in-situ} with a chemical reaction at the interface which starts as soon as the phases containing reactive molecules are in contact. Understanding and controlling the stabilization mechanisms at the interface in both ways via analysis of the kinetics and equilibrium of adsorption/desorption/reaction processes is therefore of prime interest in order to control the properties of the dispersion.

A wide body of literature is available on the adsorption/desorption mechanisms and the stabilization efficiency in blends and emulsions stabilized by pre-made surfactants \cite{Leibler1991,Bibette1996,Eastoe2000,Ravera2005,Goodwin2009,Anastasiadis2010,He2015}, and on the applicability of the reactive compatibilization \cite{Liu1992,Koning1998,Utracki2002,Macosko2005}. However, even though reactive stabilization is majoritarily used in industry for polymer blends for example, we could only find one review from 2002 dealing with the interfacial behaviour of reactive blends \cite{Litmanovich2002}. In that regard, this review is dedicated to provide new insights into the kinetics and equilibrium behaviour of interfacial tension at reactive interfaces, and to understand the implications on both the interface and the dispersion morphologies.

To do so, in Section~\ref{sec:nonreactive} we first concentrate on the definition and behaviour of interfacial tension at non-reactive interfaces, in order to compare with its behaviour at interfaces where a chemical reaction occurs in Section~\ref{sec:reactive}. Finally, in Section~\ref{sec:consequences} we review the impact of the reactive stabilization by considering the stability of the dispersion, interfacial instabilities and drop size evolution.

\section{Non-reactive interfaces}   
\label{sec:nonreactive}

In order to be able to compare the non-reactive and reactive interfaces and to allow the reader to get a broad idea on both, we begin this review with a summary of the properties of the non-reactive interfaces. First, we briefly describe in Section~\ref{Int-tens} the interfacial tension, microscopically and thermodynamically, and the molecules used to stabilize interfaces called surfactants in Section~\ref{surfactants}. Then in Section~\ref{evolsurftension} we will consider the effect of the surfactants on the interfacial tension both regarding kinetics and equilibrium. Finally, in Section~\ref{subsec:stab}, we discuss the stabilization process of emulsions and polymer blends in light of these considerations.   

\subsection{Interfacial tension between fluids}
\label{Int-tens}

\subsubsection{Microscopic origin}
\begin{figure}[h]
\begin{center}
\floatbox[{\capbeside\thisfloatsetup{capbesideposition={right,center},capbesidewidth=6cm}}]{figure}[\FBwidth]
{\caption{Scheme of an interface between two phases A and B showing the interactions between two A molecules, between two B molecules and between one A molecule and one B molecule.}\label{2.1.1}}
{\includegraphics[scale=0.7]{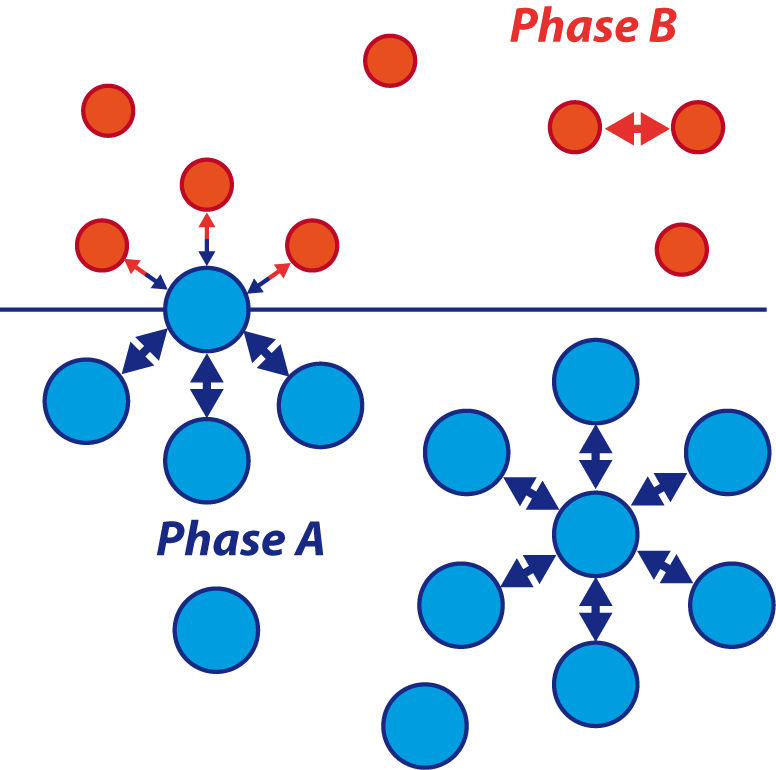}}
\end{center}
\end{figure}

In a liquid phase, molecules exert attractive forces on each other, such as Van der Walls forces or hydrogen bonding among others. These keep them close, in opposition to a gas phase. A molecule located in the bulk of an isotropic liquid interacts only with molecules that happen to be identical to it, meaning that the total force field exerted on each molecule is isotropic. In the case of a molecule at the interface with another liquid, a part of the attractive forces felt by the molecules are exerted by molecules of a different phase. Since interaction forces between molecules (Van der Walls forces, hydrogen bonding, polar interactions, etc.) are stronger between two identical molecules than between two different ones, the cohesive energy at the interface is weaker than in the bulk (Figure~\ref{2.1.1}). The interfacial tension $\gamma_{A/B}$ at the interface between two phases A and B, equivalently expressed in mN/m or mJ/m$^2$, measures this energy shortfall per unit surface area. This implies that the stronger the interactions between two identical molecules, the higher the interfacial tension at the interface. For strongly cohesive liquids such as mercury, the interfacial tension against air is $\gamma_{mercury/air}\approx500\:$mN/m \cite{deGennes2013}. Low cohesive liquids like oils present interfacial tensions of $\gamma_{oil/air}\approx20\:$mN/m. Water, because of hydrogen bonds between the molecules, has a higher interfacial tension against air $\gamma_{water/air}\approx72\:$mN/m.    

\subsubsection{Thermodynamic definition}

If it costs energy for a molecule to stay at the interface, it costs energy to create an interface. In a system composed of two phases $\alpha$ and $\beta$ and an interface $I$, we can define the internal energy $U$ as 
\begin{equation}
U=U^{\alpha}+U^{\beta}+U^I \text{,}
\end{equation}
where $U^{\alpha}$ and $U^{\beta}$ are the internal energies of both homogeneous phases and $U^I$ represents the excess internal energy arising from the presence of an interface, and can vary under transfer of mass in and out of the interface \cite{NeumannBook2010}. In this case, the change in internal energy is given by 
\begin{equation}
dU=TdS-pdV+\gamma dA+\sum_{i} \mu_idn_i\text{,}
\end{equation}
where $TdS$ is the heat transfer with $T$ the temperature and $S$ the entropy, $PdV$ and $\gamma dA$ is the mechanical work which takes into account changes in the system volume $V$ and the interfacial area $A$ respectively, $\sum_{i} \mu_idn_i$ is the chemical work with $\mu_i$ the chemical potential of the molecule $i$ and $n_i$ the number of moles of the species $i$. The chemical potential of the species $i$ is defined as $\mu_i=kTln(a_i)$ with $k$ the Boltzmann constant and $a_i$ the activity of the molecule $i$.

From the internal energy we can define the Gibbs free energy $G=U+pV-TS+\sum_{i} \mu_i n_i$ which gives 
\begin{equation}
dG=-SdT-Vdp-\gamma dA+\sum_{i} \mu_idn_i\text{.}
\end{equation}

At constant temperature $T$, pressure $p$ and number of moles $n_i$, the interfacial tension $\gamma$ is defined by 
\begin{equation}
\gamma=\frac{\delta G}{\delta A}\Big|_{T,p,n_i}
\end{equation}
and represents the increase of free energy $\delta G$ that occurs when increasing the interface area by $\delta A$. In order to minimize its energy, the system will spontaneously minimize the interface area.

\subsection{Surface active agents}
\label{surfactants}

When mixing two immiscible liquids, the energy required to create an interface is too high and the system spontaneously tends to a minimal area of contact i.e. demixing of the phases occurs. Surface active agents, usually called surfactants, present affinities for both phases in contact. They can be short molecules \cite{Narsimhan2001,Carey2010,Varade2011}, usually hydrophobic carbon chains with a hydrophilic neutral or charged end, larger molecules such as proteins \cite{Damodaran2005,Dickinson1996} or copolymers where each polymer has a strong affinity for one of the phases \cite{Aravind2004,Sundararaj1995,Leibler1988}. In this review, we will also take the particles into account in our definition of surfactants \cite{Binks2002} due to the strong similarities of interfacial behaviour \cite{Golemanov2006,Gonzenbach2006,Dickinson2010,Gauckler2006}.

Their ambivalent affinities drive these agents towards the interface where they form an energy barrier against demixing of the two phases. This barrier is the result of interactions between two liquid films in contact, also called disjoining pressure (steric and/or electrostatic repulsion) and the visco-elastic properties of the interface: when for example two oil drops in a solution of surfactants in water come close to one another, the presence of the surfactants at both interfaces in contact prevents coalescence \cite{Hunter2008,Exerowa2003,Alargova2004}.  

\subsection{Evolution of interfacial tension in the presence of surface active agents}          
\label{evolsurftension}

\begin{figure}
\begin{center}
\includegraphics[scale=0.3]{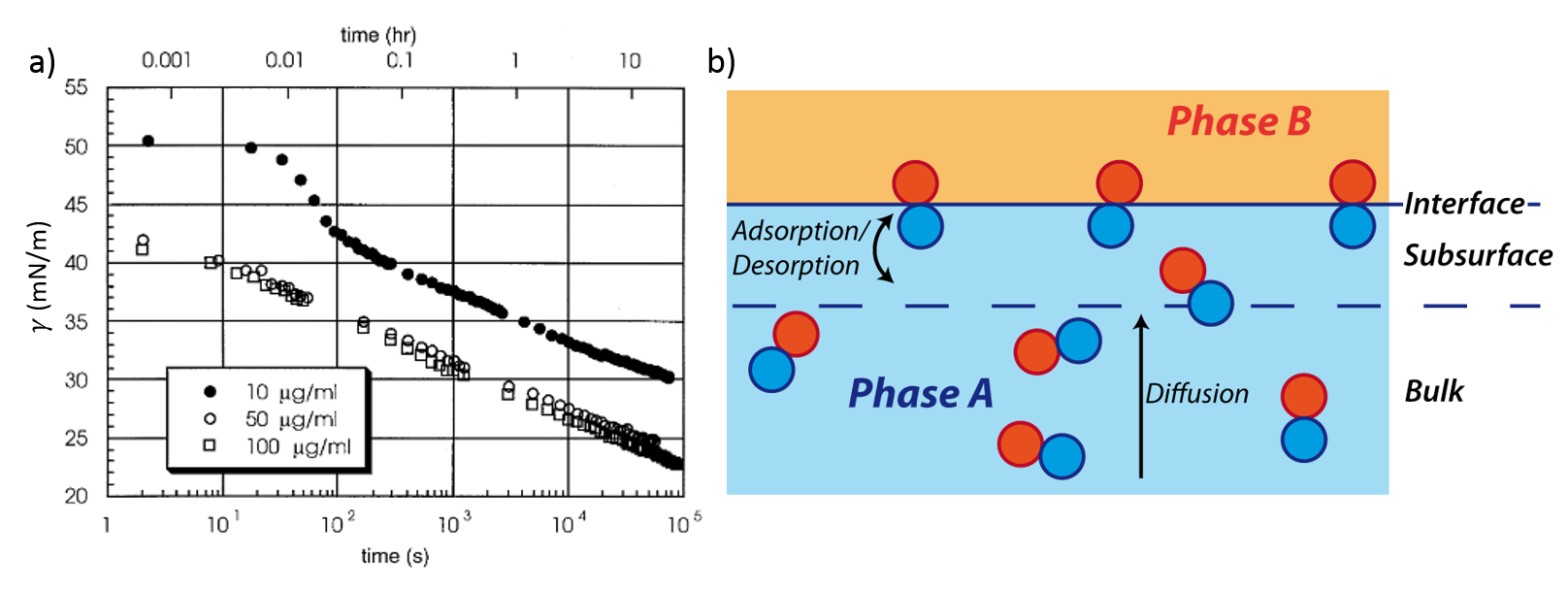}
\caption{a) Dynamic interfacial tension of ovalbumin at the interface between heptane and an aqueous buffer (pH 7.1, 100 mM sodium phosphate) interface at different bulk concentration of ovalbumin (from \protect\cite{Beverung1999} with permissions). b) Scheme of the differentiation between bulk, subsurface and interface.}
\label{2.3}
\end{center}
\end{figure}

Here, we will discuss only the case where the surfactants have a finite solubility in one of the two liquids only. As just mentioned, the primary role of surfactants is to stabilize an emulsion of two immiscible liquids. Their presence however also has an impact on the value of interfacial tension. Figure~\ref{2.3} shows a typical evolution of interfacial tension  with time just after the creation of a fresh interface, in this example between heptane and an aqueous buffer solution for different bulk concentrations of ovalbumine \cite{Beverung1999}. The interfacial tension measured at $t=0$ s is naturally the value measured between the two phases without any surfactants $\gamma_0$. It then decreases with different characteristic times $\tau$ depending on the type and concentration of the added surfactants, to finally reach an equilibrium value $\gamma_{eq}$. This overall relaxation behaviour is observed for all types of surfactants, and only the values of $\gamma_0$, $\tau$ and $\gamma_{eq}$ vary between different systems. The shape of the relaxation curve depends on the type of surfactant and can be quite complex, for example, when energy barriers of adsorption are present or when the surfactant has a finite solubility in both phases.

Initially, when a fresh interface is created, the surface concentration $\Gamma$ (in units/m$^2$) of surfactants is negligible. This creates a flux of surfactants from the bulk to first the subsurface and then the interface until an equilibrium between adsorption and desorption is reached (Figure~\ref{2.3}). Depending on the surfactant used, two models are discussed in Section~\ref{subsubsec:kinetic}.

The first model is the so-called \textit{diffusion controlled} model \cite{Ward1946}. It makes the assumption that once the surfactant has diffused from the bulk to the subsurface it is directly adsorbed at the interface. The limiting time-scale is the diffusion time from the bulk to the subsurface. 

The other model is the \textit{mixed kinetic-diffusion} model. Here, an energetic barrier is assumed for the surfactant to go from the subsurface to the interface \cite{Blair1948}, which becomes the limiting phenomenon in the kinetics of absorption at the interface. It can be attributed to the fact that there are less and less vacant sites at the interface, making it more difficult for new surfactants to adsorb with time. This can also apply when dealing with large molecules that can cause steric repulsions or need to have the right orientation to absorb, and ionic surfactants that cause electrostatic repulsions.

The equilibrium value of interfacial tension, and the kinetic models both at the very beginning of the evolution of interfacial tension ($t\rightarrow 0$) and at the equilibrium phase ($t\rightarrow \infty$) can be analysed separately.

\subsubsection{Equilibrium interfacial tension and CMC}
\label{eqGamma}

At an interface composed of solvent molecules of both phases and a surfactant (assumed to have a high solubility in one phase and a low solubility in the other), the position of the interface is conveniently chosen so that $\Gamma_{solvent}=0$. In the dilute solution approximation (when the activity $a$ can be approximated by the bulk concentration of surfactant $c$, thus $d\mu$ can be approximated by $kTd$ln($c$)), the value of equilibrium interfacial tension $\gamma$ is linked to the equilibrium surface excess of surfactants $\Gamma$ by the Gibbs adsorption isotherm
\begin{equation}\label{Gibbs}
\Gamma=-\frac{1}{mRT}\Big(\frac{\delta \gamma}{\delta\text{ln}c}\Big)_{T,V,n_i}\text{,}
\end{equation}
where $R$ is the gas constant, and $m$ depends on the nature of the surfactant. For non-ionic surfactants, $m=1$ is in very good agreement with several experimental studies \cite{Cooke1996,Lu1993,Lu-Li1993,Lu1997,Eastoe1997,Li2013}, as well as $m=2$ for 1:1 ionic surfactants\footnote{for 1:1 ionic surfactants, the charged group of the surfactant and the counterion have opposite charges ($\pm1$), where for 2:1 ionic surfactants the hydrophilic group and the counterion have charges of $\pm2$ and $\mp1$, i.e. the global molecule is non neutral.} assuming electrical neutrality of the interface \cite{Tajima1970,Penfold1996}. For 2:1 surfactants, however, discrepancies in the experimental results motivated the use of an adjustable parameter $\alpha$ caused by ion impurities at the interface \cite{Hall1975,Penfold1996}, and led to $m=2-\alpha$.

\begin{figure}[H]
\begin{center}
\includegraphics[scale=0.3]{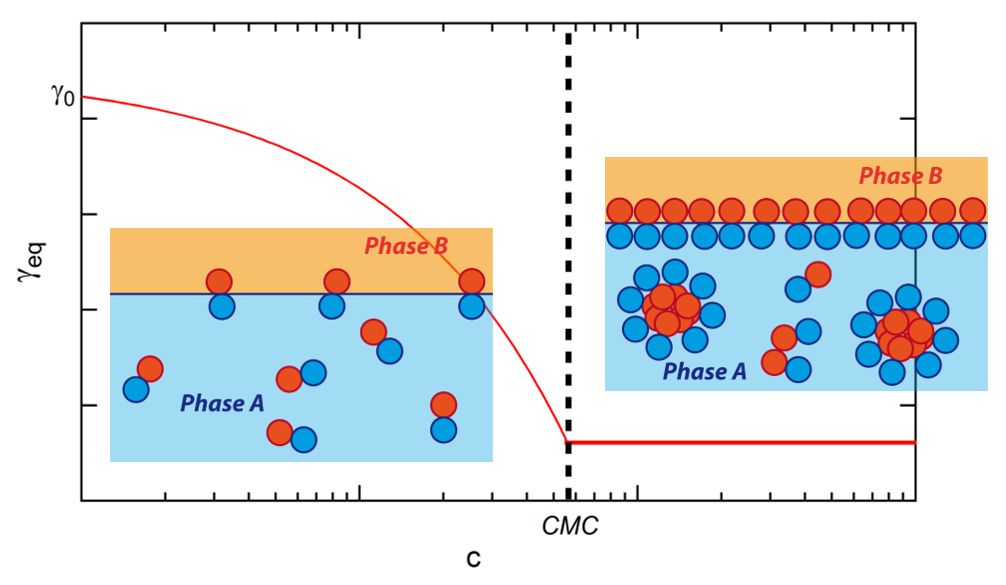}
\caption{Scheme of a generic evolution of the equilibrium interfacial tension $\gamma_{eq}$ against the bulk concentration of surfactants $c$.}
\label{2.3.1}
\end{center}
\end{figure}

Looking at the value of interfacial tension at equilibrium against the bulk concentration allows to obtain the \textit{critical micellar concentration}, or $CMC$, which is the bulk concentration above which the surfactants start to create micelles in the bulk. For $c>CMC$, the equilibrium interfacial tension $\gamma_{eq}$ will be independent of $c$. This is shown by $\gamma_{eq}(c)$ curves (Figure~\ref{2.3.1}), where for $c<CMC$ the value of $\gamma_{eq}$ decreases with $c$, and then reaches a plateau value. The $CMC$ is defined by the crossover between the plateau and the decreasing curve.

The Gibbs isotherm is a way of measuring indirectly the equilibrium surface concentration of surfactants $\Gamma_{eq}$ with varying $c$.

\subsubsection{Kinetic description} 
\label{subsubsec:kinetic}

\paragraph{Diffusion-controlled interfaces}\hspace{0.5cm}

A widely used mathematical model to describe the adsorption processes of surfactants at an interface where the limiting time-scale is controlled by diffusion is the model proposed by Ward and Tordai in 1946 \cite{Ward1946}. This model takes into account the progressive filling of the interface by surfactants, meaning that it becomes more and more difficult for them to adsorb because of interactions with the already adsorbed surfactants, and consequently allows back diffusion, i.e. desorption from the interface. This equation however cannot be solved analytically, which is why Miller, Fainerman and Makievski \cite{Fainerman1994} derived asymptotic solutions to this equation to describe the kinetics of the interfacial tension decay. These solutions look at the behaviour of the interfacial tension at $t\rightarrow 0$ and $t\rightarrow \infty$. Here, we assume for simplicity that the surfactant is insoluble in one of the phases. 

Initially, there are almost no surfactants at the interface (i.e. $\gamma \rightarrow \gamma_0$), so they will not desorb from the interface. This, in the case of an ideal surface layer, gives: 
\begin{equation}
\gamma_{t\rightarrow 0}=\gamma_0-2nRTc\sqrt{\frac{Dt}{\pi}}\text{,}
\end{equation}
where $D$ is the diffusion coefficient of the surfactant in the phase it is dissolved in.

When $t\rightarrow \infty$, the subsurface concentration tends to the bulk concentration and allows, along with the Gibbs equation, to get: 
\begin{equation}\label{eq7}
\gamma_{t\rightarrow \infty}=\gamma_{eq}+nRTc\sqrt{\frac{\pi}{Dt}}\text{.}
\end{equation}

This diffusion-controlled model generally works very well for $t\rightarrow 0$ and at low $c_0$ . 

\paragraph{Mixed kinetic-diffusion controlled adsorption}\hspace{0.5cm}

Equation (\ref{eq7}) on the other hand was shown to be ineffective to capture the behaviour of the interfacial tension at longer times by several experimental studies\cite{Lin1995,Lin1996,Chang1998,Eastoe1996,Eastoe1997,Eastoe1998}. The solution to this problem was theoretically implemented by Barret \textit{et al.} \cite{Baret1968} and Liggieri \textit{et al.} \cite{Ravera1993,Liggieri1996} and is called the mixed kinetic-diffusion controlled adsorption.

This model relies on the assumption that once a surfactant has diffused from the bulk to the subsurface, it may not instantly adsorb at the interface, for different possible reasons, mainly related to steric or electrostatic repulsions and need of the right orientation (in the case of long chain molecules i.e. proteins and polymers) \cite{Alexander1963,Graham1979,Cho1997,Cornec1999,Mahato2009}. All of this can be understood as the existence of an activation energy $\epsilon_a$ that the surfactant has to overcome to itself go from the subsurface to the interface. It induces a renormalized diffusion coefficient \begin{equation}D^*=D \text{exp}(-\epsilon_a/RT)\text{,}\end{equation} theoretically predicted \cite{Baret1968,Ravera1993,Liggieri1996}, and confirmed experimentally \cite{Lin1995,Lin1996,Chang1998,Eastoe1996,Eastoe1997,Eastoe1998}. By using $D^*$, it is possible to consider this a diffusion problem again and to change the Ward and Tordai equation accordingly. Most systems are well described by the simple diffusion model at short times and switch to the mixed kinetic-diffusion model with increasing time, even for $c<CMC$. % 

\subsection{Application to stabilization processes}\label{subsec:stab}

Surfactants are used in the making of emulsions and polymer blends to stabilize them against coalescence (rupture of the film between two drops) and Ostwald ripening (diffusion of molecules from small to large drops). The diffusion time $\tau$ from the bulk towards the interface and its dependence on the surfactant size, and desorption from the interface then have important consequences on the stability of the dispersion.

When dealing with phases with low viscosities and low molecular weight surfactants, this stabilization process proves to be efficient \cite{Du2003,Dickinson2004,Schroder2016,Bibette1996,Uhlig2016}, thanks to the fact that small surfactants can easily diffuse in a low-viscosity medium. In these conditions, the surfactant will rapidly go from the bulk to the interface to stabilize the emulsion. The life-time of the emulsions and blends depends on the surfactants ability to create repulsive interactions against one another upon the approach of two interfaces (disjoining pressure) \cite{Stubenrauch2003}, and on the visco-elastic properties which they confer to the interface \cite{Langevin2012}.

Heterogeneous polymer blends, which are dispersions of two immiscible polymer melts \cite{Koning1998}, are more difficult to stabilize than the cases discussed above for several reasons. As mentioned before, surfactants come under different configurations, from small amphiphilic molecules (neutral or ionic), to large molecules such as proteins and polymers. If emulsions of two simple liquids can be stabilized with any of those, in the case of complex liquids such as polymer-polymer interfaces, it is more efficient to use surfactants in the form of block or graft copolymers. These are composed of segments that present complementary affinities towards the two phases of the blend \cite{Fayt1981,Jayabalan1985,Schwarz1989,Xu1998,Chen2004,Auschra1993,Kim1993}, to mimic the amphiphilic nature of low molecular weight surfactants. Of course these molecules are large, and as a result they diffuse slowly which causes slow adsorption kinetics at the interface. This, associated to the fact that the $CMC$ is easily exceeded with large molecules, results in poor stabilization of the polymer blend \cite{Koning1998,SundararajMacosko1995}. 

Another drawback of the use of surfactants in the stabilization process is the possible presence of micelles in the phase they were first added for any concentration $c$ below or above the $CMC$, especially in the case of block-copolymers. If the lifetime of the micelle is longer than the time for the interface to reach equilibrium, i.e. if the thermodynamic barrier of breaking a micelle is high, then less molecules adsorb at the interface and participate to the stabilization process \cite{Eastoe2000}.

To avoid the problems caused by the use of pre-made surfactants, reactive compatibilization, which consists of provoking a chemical reaction at the interface only that creates a surfactant in-situ, has been largely used in immiscible polymer blending \cite{Boucher1996,Boucher1997,Legros1997,Laurens2004,Bonnet2012,Liu1992,Legros1994,Pesneau2001,Barraud2012,Koulic2004,Marosi2004,Macosko2005,Giustiniani2016}.

\section{Evolution of interfacial tension at reactive interfaces}
\label{sec:reactive}

Stabilization of immiscible liquid dispersions is a subject of great interest in commercial applications. The use of pre-made surfactants to stabilize the drops has proven not to be effective enough for viscous systems because of slow adsorption kinetics and poor adhesion at the interface. Indeed, in order to stabilise efficiently the drops in a dispersion and since the dynamics of adsorption are so slow and impaired by the creation of micelles, the block-copolymers need to be irreversibly adsorbed at the interface. This is possible only if they are insoluble in either phase, because of their size for example, which means that they cannot be initially dispersed in one of the phases. As, if they cannot be dispersed, they cannot adsorb to the interface, reactive compatibilization of immiscible dispersions is used, i.e. the stabilising agent is created in-situ, at the interface. While this approach has  mostly been developed for polymer blends, it is now increasingly used for emulsions.

Here, in Section~\ref{subsec:methods} we summarize the principal methods of compatibilizing polymer blends and emulsions by a chemical reaction. Then we study in Section~\ref{subsec:eq} the equilibrium of interfacial tension at reactive interfaces and try a comparison with the non-reactive case studied previously in Section~\ref{eqGamma}. Finally, in Section~\ref{subsec:kin} we focus on the kinetics of evolution of interfacial tension and surface concentration in reactive systems. 

\subsection{Methods of reactive compatibilization}
\label{subsec:methods}

\begin{figure}
\begin{center}
\includegraphics[scale=0.6]{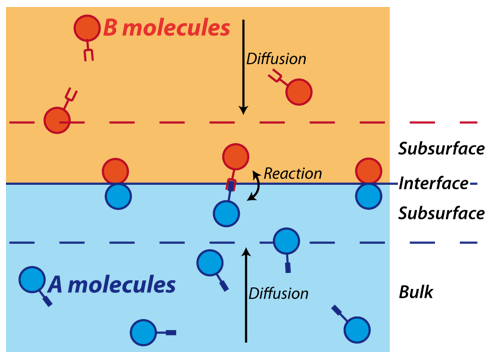}
\caption{Scheme of a reactive interface between two phases A and B containing reactive molecules, with respective concentrations $c_A$ and $c_B$.}
\label{3.1}
\end{center}
\end{figure}

Compatibilization of interfaces by a chemical reaction occurs by functionalizing the molecules of the different phases with complementary functional groups (Figure~\ref{3.1}) \cite{Boucher1996,Boucher1997,Legros1997,Laurens2004,Chi2007,Bonnet2012,Barraud2012}, sometimes with the help of a precursor \cite{Legros1994,Pesneau2001,Koulic2004,Marosi2004,Giustiniani2016}. As soon as the two liquids are in contact, the chemical reaction starts and so does the stabilization. As functionalized molecules are much more expensive than not functionalized one, usually only a certain percentage of reactive molecules is added in each phase. The percentage needed to stabilize the blends depends on the size of the molecules, and the reaction kinetics.

These methods of reactive compatibilization are to be adapted to the two polymers used in the blends. If the methods of reaction differ, the result is the same for all of them once the protocol is optimized: the blend is stabilized against demixing.

Depending on the reaction occurring at the interface, different molecules can be created at the interface. In most cases \cite{Boucher1996,Boucher1997,Legros1997,Laurens2004,Chi2007,Bonnet2012,Barraud2012,Legros1994,Pesneau2001,Koulic2004,Marosi2004,Giustiniani2016}, the reaction creates in-situ block or graft copolymers which act as surfactants at the interface. 

\subsection{Equilibrium interfacial tension}
\label{subsec:eq}

Experiments show that the relaxation of the interfacial tension of reactive interfaces exhibits a strikingly similar behaviour to that of non-reactive interfaces \cite{Chi2007,Zhaohui2001,Powell2016,Kim2000,Cho1996,Giustiniani2016}. All curves start at a value $\gamma_0$, which is the value of interfacial tension between the two non-reactive phases with no added surfactants. The surface tension $\gamma$ then decreases with a characteristic relaxation time that depends on the bulk concentration of reactive polymers until it reaches an equilibrium value $\gamma_{eq}$. Can we then compare the behaviour of reactive and non-reactive interfaces? To answer this question, we will look first at the effect of the copolymer configuration at the interface, and then try to determine if the Gibbs equation (Equation (\ref{Gibbs}) Section~\ref{eqGamma}), used for non-reactive systems, is also applicable in the case of in-situ formed surfactants.

\begin{figure}
\begin{center}
\includegraphics[scale=0.3]{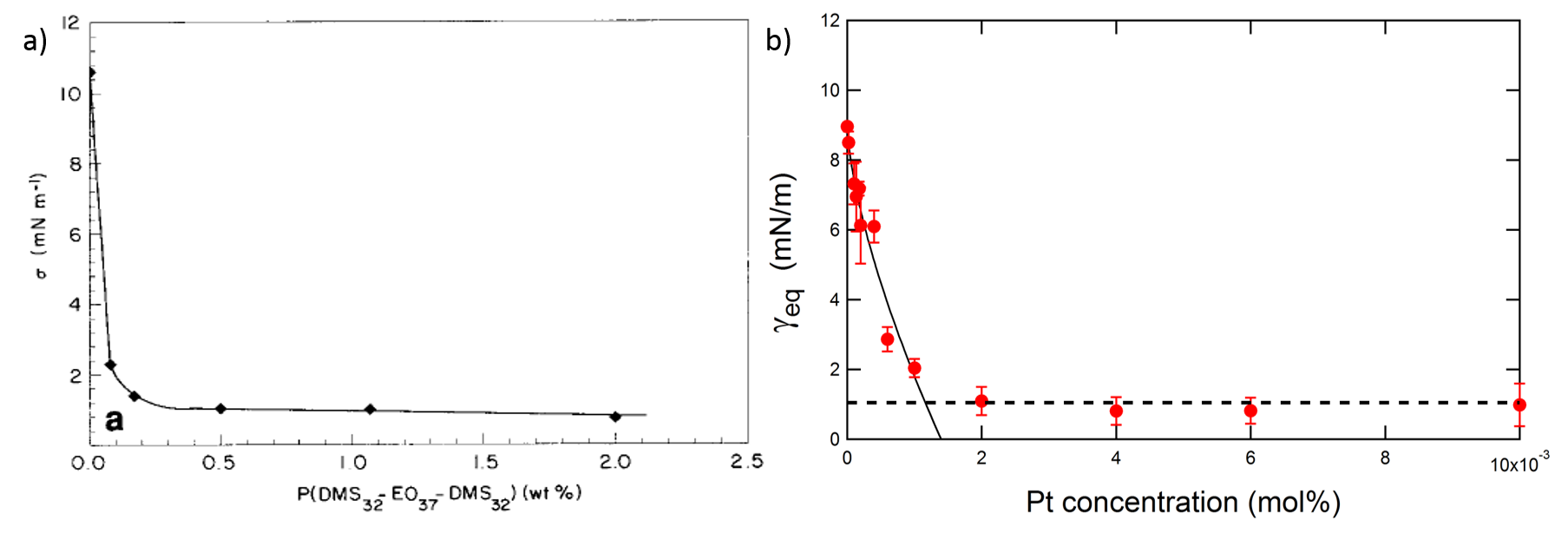}
\caption{a) Interfacial tension (in mN/m) between PEO 35 ($M_n=32,000$ g/mol) and PDMS 100 ($M_w=177,000$ g/mol) for different concentrations of pre-made P(DMS$_32$-EO$_37$-DMS$_32$) copolymers (in wt\%) in the PEO phase (from \protect\cite{Wagner1993} with permissions). b) Interfacial tension between PEG 400 ($M_w=400$ g/mol) and reactive PDMS ($M_w=2,000$ g/mol) for different concentrations in platinum Pt (in mol\%) in the PEG phase which allows the reaction between the PEG and the reactive PDMS (adapted from \protect\cite{Giustiniani2016} with permissions).}
\label{3.2}
\end{center}
\end{figure}

Comparing the properties for different pairs of immiscible polymers stabilized by either pre-made copolymers or in-situ formed copolymers shows that the equilibrium interfacial values are identical with both stabilization routes. For example, Wagner \textit{et al.} \cite{Wagner1993} measured the interfacial tension between PDMS and PEG with pre-made PDMS-PEG-PDMS and found equilibrium values close to the ones which Giustiniani \textit{et al.} \cite{Giustiniani2016} found for PDMS and PEG stabilized by in-situ compatibilization (Figure~\ref{3.2}). As the configuration of the copolymers, in addition to their chemical composition, has an impact on the equilibrium value of interfacial tension $\gamma_{eq}$ \cite{Lyatskaya1995,DiLorenzo1997,Cho1996}, the comparisons between the reactive and non-reactive cases is not straightforward. Indeed, while in non-reactive systems, the control over the chemical architecture of the surfactant is clear, in reactive systems the obtained molecules can have several reactive groups in their architecture which leads to more polydispersity in the surfactants at the interface. This also has consequences on the stabilization efficiency and the adhesion between the phases after solidification \cite{Edgecombe1998,Creton2001}. 

For non-reactive interfaces, the Gibbs adsorption isotherm (Equation (\ref{Gibbs}) Section~\ref{eqGamma}) links the interfacial tension to the surface coverage of surfactants. The principal hypothesis that allows to derive the Gibbs equation as it is given in Section~\ref{sec:nonreactive} is the insignificance of the variation of the solvent's chemical potential. In the case of small surfactants in water for example, the Gibbs-Duhem equation gives $X_{water}d\mu_{water}=-X_{surf}d\mu_{surf}$ where $X$ is the molar fraction and $\mu$ the chemical potential of the water and the surfactants. This implies that $d\mu_{water}\ll d\mu_{surf}$. Added to the fact that the solution has to be considered dilute, the Gibbs equation stands. To analyse their measures of surface tension relaxation at a reactive interface, Chi \textit{et al.} also used the Gibbs equation \cite{Chi2007}. They obtained access to the equilibrium surface concentration $\Gamma^\infty$ , and proved the Volmer model of adsorption kinetics \cite{Volmer1925} \footnote{The volmer model derives from the Gibbs equation and accounts for the non-ideal non-localised adsorption at the interface and the finite size of the surfactant molecules.} to be effective in capturing the $\gamma$-$\Gamma$ curve. They state that their analysis relies on the fact that less than 3\% of the initially added reactants are used in the process, so the bulk concentration of reactant at equilibrium is approximately equal to its initial value.

In most cases, the surface concentration of copolymers $\Gamma$ has been shown to be measurable directly and independently of the interfacial tension $\gamma$, either by XPS analysis of the interface \cite{Oyama2001,Zhaohui2001,Jones2003,Guegan1994,Yin2001,Yu2005}, gel permeation chromatography (GPC) \cite{Jeon2004} or dynamic secondary ion mass spectrometry (DSIMS) with a deuterium labelled reactant \cite{Kim2005}. The interfacial tension is measured by the pendant drop method \cite{Chi2007,Giustiniani2016} or the Neumann triangle method \cite{Kim2000,Tanaka2016}.

The overall relaxation and equilibrium value of interfacial tension in reactive systems is comparable to non-reactive systems stabilized with pre-made surfactants. The same adsorption isotherms seem to be applicable to reactive systems only in specific conditions, namely large excess of reactants.

\subsection{Kinetics of evolution of interfacial tension}   
\label{subsec:kin}

In order to form in-situ surfactants at the interface, the reactive molecules in both phases in contact have the same kinetic limitations as pre-made surfactants, namely diffusion towards the interface, and existence of an energetic barrier due to steric or electrostatic repulsions. The latter can be reduced to a diffusion limited problem with an exponentially decreasing diffusion coefficient, as stated in Section~\ref{subsubsec:kinetic}. But in reactive systems, to these energy barriers has to be added the kinetic of the reaction itself, which requires of each reactive molecule to i) find a reaction partner and ii) react with it. In the presence of a reaction at the interface, the kinetics of the evolution of the surface concentration of surfactants $\Gamma$ follows the rate equation \cite{Fredrickson1996}
\begin{equation}
\frac{d\Gamma}{dt}=k\rho_{A}(t)\rho_{B}(t)\text{,}
\label{eq13}
\end{equation}
where $k$ is the reaction coefficient, and $\rho_{A}$ and $\rho_{B}$ the number density of reactive chains A and B in the vicinity of the interface. The densities $\rho_{A}$ and $\rho_{B}$ vary with time according to diffusion and reaction at the interface. We call $\tau_{d}$ and $\tau_{r}$ the characteristic diffusion and reaction times respectively. If $\tau_{r}\ll\tau_{d}$, i.e. the reaction takes place instantly, then the only limiting factor is the diffusion of the reactive molecules towards the interface linked with the probability of meeting a reaction partner. If, on the contrary, the reaction is slow compared to the diffusion, $\tau_{r}\gg\tau_{d}$, then the reaction is the limiting factor of the evolution of interfacial tension. 

\subsubsection{Diffusion-controlled interfaces}

The simplest case here is when $\tau_{r}\ll\tau_{d}$ and the system is diffusion-limited. Assuming for simplicity that the initial concentrations of reactive chains A and B in each phase is low and equal for both A and B ($\rho_0=\rho_A=\rho_B$), and that A and B have the same degree of polymerization($N=N_A=N_B$), Fredrickson and Milner \cite{Fredrickson1996} showed that the kinetics of the creation of copolymers at the interface exhibits three regimes, with different associated characteristic times. They define the times $\tau_\rho$ as the characteristic time for the number density of reactive ends $\rho$ in the interfacial region to decay, and $\tau_\Gamma$ as the characteristic time of the evolution of copolymer coverage of the interface. They give the relations 

\begin{equation}\tau_\rho=\tau\frac{\text{ln}^2N}{(\rho_0R^3)^2} \qquad\text{and}\qquad\tau_\Gamma=\frac{(\Gamma^{\ast})^2}{D_0\rho_0^2} \text{,}
\end{equation} 
with $\tau$ the disentanglement time, also called the terminal relaxation time, $\Gamma^{\ast}$ the copolymer coverage of the interface at which the chemical potential barrier is of order $k_BT$ (with $k_B$ the Boltzmann constant), and $D_0$ the center-of-mass diffusion coefficient of a reactive chain in the bulk. When $N<N_e$, $N_e$ being the entanglement threshold, $\tau \sim N^2$ is the Rouse time. For entangled chains ($N>N_e$), $\tau \sim N^3$ is the reptation time.

In the short time regime ($0<t<\tau_\rho$), there are already reactive chains at the interface, so the kinetic of evolution of $\Gamma$ is determined by the reaction rate and the number densities $\rho_{A}$ and $\rho_{B}$ are approximately constants. Equation (\ref{eq13}) becomes 
\begin{equation}\frac{d\Gamma}{dt}\approx k' \qquad\text{which gives}\qquad \Gamma  \sim t \qquad \text{for} \quad t<\tau_\rho \text{.}
\label{eq11}
\end{equation} 

If the reaction is fast enough, the densities $\rho_{A}$ and $\rho_{B}$ start to dramatically decrease for $t\sim\tau_\rho$, and a depletion hole is created near the interface. The copolymer coverage of the interface is, in this intermediate regime ($\tau_\rho<t<\tau_\Gamma$), diffusion controlled, and the evolution of surface concentration is 
\begin{equation}
\Gamma  \sim t^{1/2} \qquad \text{for} \quad \tau_\rho<t<\tau_\Gamma \text{.}
\end{equation} 

When $t>\tau_\Gamma$, the copolymer coverage saturates starting from a value of the order $\Gamma^{\ast}\sim b^{-2}N^{-1/2}$ where $b$ is the statistical segment length. Looking at the evolution of $\Gamma$ during this final regime gives out 
\begin{equation}
\Gamma  \sim (\text{ln}t)^{1/2} \qquad \text{for} \quad t>\tau_\Gamma \text{.}
\end{equation}  

It is worth paying special attention here to the different assumptions of this theory. First, the reactive polymers are assumed to have the same size. Yet, experimental studies often involve reactions between two polymers of different sizes, which implies different mobilities for each species and lead to qualitative changes in the results. Also, this study assumes very low initial concentrations of reactive species in both phases. Monte Carlo simulations carried out by M\"{u}ller \cite{Muller1997} showed that this theory for diffusion-controlled reactions at interfaces given by Fredrickson and Milner was indeed verified only for very low initial concentrations in reactive species. But most importantly, here it is assumed that the chemical reaction is fast. However, this is not the case in most reactive blending processes, as pointed out by O'Shaughnessy and Vavylonis \cite{OShaughnessy1999}. This could explain why it is difficult to find experimental studies in agreement with the diffusion-controlled theory. Schulze \textit{et al.} \cite{Schulze2000} compared their experimental results with the diffusion-limited theory of Fredrickson and Milner but found no agreement.

\subsubsection{Reaction-limited systems} 

By comparing the estimated diffusion times of the reactive molecules in the bulk, from their size and the viscosity of the medium they are in, to the characteristic time of evolution of interfacial tension found experimentally, one can already have an idea of the kinetics of the reaction. For a system of PDMS-NH$_2$ in PDMS, with molecular weights ranging from $M_n=27,000$ to $62,700$ g/mol, and by considering that only the PDMS-NH$_2$ in a layer of thickness $h$ actually play a role in the reactive stabilization, Chi \textit{et al.}\cite{Chi2007} found a range of diffusion times from 20 s for the lowest concentration of reactive PDMS to 1.6 s for the highest concentration. Since the characteristic time of evolution of interfacial tension for their system is at minimum of the order of 2-5 min, they concluded that the system was reaction-limited. In that case, $\tau_{r}\gg\tau_{d}$ and the evolution of interfacial tension cannot be described as a diffusion problem with a varying coefficient of diffusion as for non-reactive interfaces. The evolution of interfacial tension depends on the reaction kinetics, which depends on the rate of the reaction taking place at the interface i.e. on the reactivity of the functional groups \cite{Orr2001}. 

Different approaches have emerged concerning the kinetics of evolution of interfacial tension of reaction-controlled systems. O'Shaughnessy and Vavylonis \cite{OShaughnessy1999} state, using a mean-field theory approach, that for small reactivities between the functional molecules A and B, at short time scales and for a flat interface, the concentrations of reactive molecules close to the interface is constant, as in the short time scales regimes for diffusion-controlled reactions, meaning that equation (\ref{eq11}) stands and $\Gamma \sim t$. They explain that if both initial concentrations are approximately equal, this persists until the surface is crowded by A-B copolymers, but assuming that $c_{B,0}<c_{A,0}$, a depletion of B in the vicinity of the interfaces forces a crossover to a first-order diffusion-controlled kinetics. Berezkin \textit{et al.} \cite{Berezkin2013} used this theory at a flat interface to study the kinetic behaviour at a curved interface (drop) using dissipative particle dynamics modelisations, and showed that due to the curvature the size ratio between A and B played a crucial role in the surface coverage, though the same kinetic laws still applied.

Oyama \textit{et al.} \cite{Oyama2001} applied a general treatment, commonly used in surface science for reactions at the gas/solid interface, to a liquid/liquid interface, and compared their finding with both their own experiments and experiments of other groups. The comparison with a gas/solid interface implies that the reaction rate is proportional to the number of vacant reactive sites at the interface. As both Fredrickson and Milner \cite{Fredrickson1996} and O'Shaughnessy and Vavylonis \cite{OShaughnessy1999}, they state that the limiting factor can either be the saturation of the interface (at a surface concentration $\Gamma^{\ast}$), or the depletion of either A or B. They take into account different considerations concerning the initial concentrations of A and B which lead to a change in the reaction kinetics. Assuming that there is no reverse reaction, that the formed interface is flat and two-dimensional between the polymers and that the copolymers formed at the interface cannot desorb, they give the rate equation for an interfacial reaction within a limited sub-surface
\begin{equation}
\frac{d\Gamma}{dt}=k(c_A-\Gamma)(c_B-\Gamma)(\Gamma^\infty-\Gamma)\text{,}
\end{equation}
where $k$ is the reaction constant, $c_A$ and $c_B$ the number of molecules A and B respectively per area of sub-surface at time $t$ (within the distances $\sqrt{2D_{A,B}t_\infty}$), and $t_\infty$ is the reaction time needed to reach the final surfactant concentration at the interface $\Gamma^{\infty}$. In this equation, $(\Gamma^\infty-\Gamma)$ represents the number of available reaction sites at time $t$. Two cases seem to be the most useful for interface stabilization, and they assume in both that A is always in excess ($c_{A} \gg \Gamma$ at all $t$).

They show that, if B is also in excess, the reaction goes on until the interface is saturated and $\Gamma^{\infty}=\Gamma^{\ast}$, where $\Gamma^{\ast}$ is the concentration of surfactants when the interface is saturated. In that case, $c_{B} \gg \Gamma$ at all $t$ and the rate equation becomes 
\begin{equation}
\frac{d\Gamma}{dt}=kc_{A,0}c_{B,0}(\Gamma^\infty-\Gamma(t))\text{.}
\end{equation} The reaction is of order\footnote{The order of a reaction is defined as the exponent to which its concentration term in the rate equation is raised.} 1 and its solutions are in the form of exp$(-t)$. The addition of the term $-k_r\Gamma(t)$ in the equation allows to take into account the reverse reaction, $k_r$ being its reaction constant \cite{Chi2007}. The solutions in that case are in the same form but the pre-factors change.

If B is not in excess, then the reaction stops when there is a depletion of B in the vicinity of the interface, and $\Gamma^{\infty}<\Gamma^{\ast}$. In that case, the rate equation becomes: 
\begin{equation}
\frac{d\Gamma}{dt}=kc_{A,0}(\Gamma^\infty-\Gamma(t))\text{.}
\end{equation}
Here also, the reaction is of order 1, and the solutions are in the form of exp$(-t)$.  

Both these equations give access to the reaction constant $k$ through the slope of the curve at short times $\frac{d\Gamma}{dt}(t \rightarrow 0)=kc_{A,0}c_{B,0}\Gamma_\infty$ or $\frac{d\Gamma}{dt}(t \rightarrow 0)=kc_{A,0}\Gamma_\infty$.

\begin{figure}
\begin{center}
\includegraphics[scale=0.3]{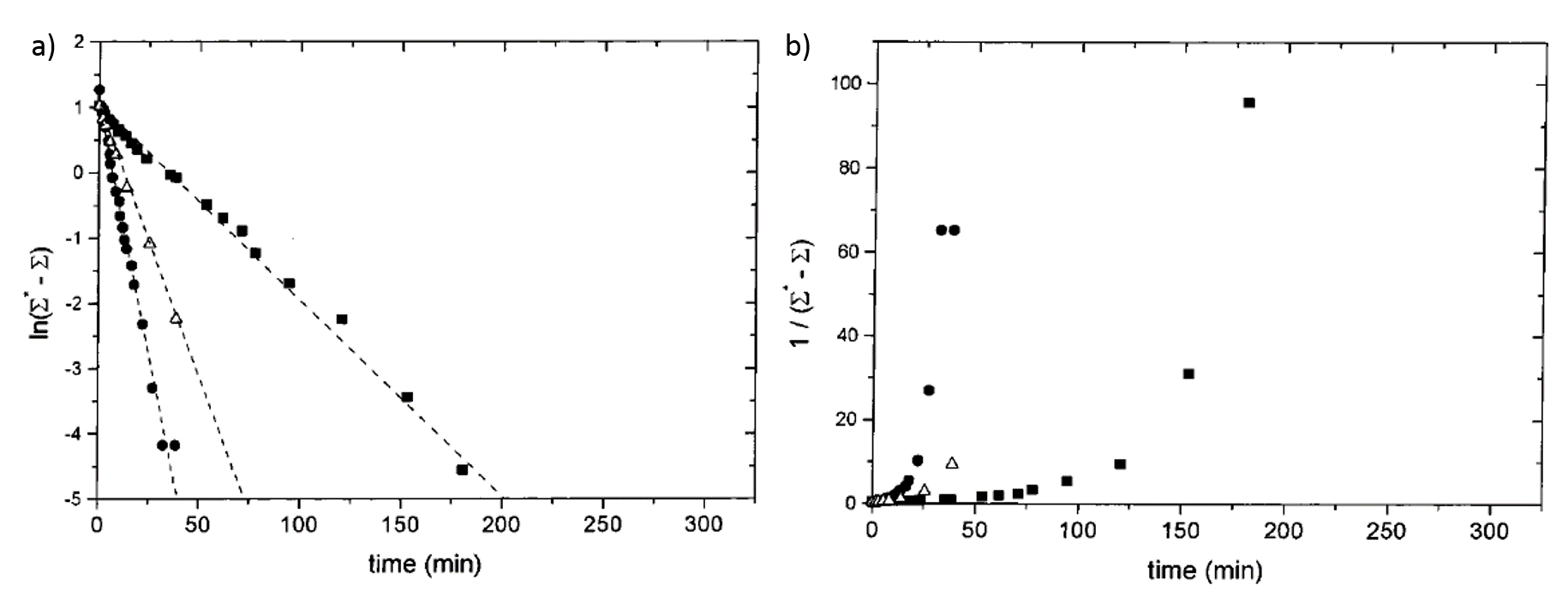}
\caption{Reaction at the interface between SMA (styrene-maleic anhydride copolymer) and ATBA (amine terminated butadiene-acrylo-nitrile copolymer) (data from \protect\cite{Scott1994}, analysis from \protect\cite{Oyama2001}, with permissions). a) First-order plot. b) Second-order plot.  $\blacksquare$ reaction at 140 $^{\circ}$C, $\triangle$ reaction at 150 $^{\circ}$C, $\bullet$ reaction at 160 $^{\circ}$C.}
\label{3.3.2}
\end{center}
\end{figure}

Experimentally, the first order dynamics has been confirmed for different systems. Oyama \textit{et al.} \cite{Oyama2001} analysed the data of Scott \textit{et al.} \cite{Scott1994} trying both a first and second-order dynamics (where $1/(\Gamma_{eq}-\Gamma)\sim t$) which showed that only the first-order was in agreement with the experiments (Figure~\ref{3.3.2}). They also confronted their theory to the one of Kramer \cite{Kramer1995} who assumed a reaction-controlled dynamics but with a reaction of order 2 and showed that the first-order dynamics fitted better the data. Kim \textit{et al.} \cite{Kim2003} compared their experimental data of the kinetics of evolution of the viscosity of the blend with fits of the first and second-order kinetics equations, and observed that both equations could fit their data in the initial stage, but only the first-order was capable of capturing the evolution throughout the entire data set. Chi \textit{et al.} \cite{Chi2007} observed a very good agreement of the first-order kinetic for small reactant concentrations at all times, but showed that for higher concentrations, the model of Oyama \textit{et al.} \cite{Oyama2001} only captured the initial growth of the surface concentration. They attributed this to chain-chain interactions becoming progressively rate limiting as the interface approaches saturation. Jiao \textit{et al.} \cite{JiaoKramer1999} were able to fit the curve of surface copolymer coverage against time between functionalized polystyrene and poly(styrene-r-maleic  anhydride) by a function of the form $\Gamma \sim (1-\text{exp}(-t/\tau))$ for the entire time-scale of the experiment, in agreement with the theory developed above. 

Jeon \textit{et al.}\cite{Jeon2004} showed that the functional group location along a polymer chain has a strong impact on the coupling reaction. Due to steric effects, a bulk reaction with end-functionalized polymers is sensibly faster than with mid-functionalized polymers. This effect is enhanced at the interface.

The kinetic behaviour of reactive liquid/liquid interfaces seems more complicated to analyse than their equilibrium behaviour. Indeed, the balance between the reaction time between the pair of functionalized molecules, and their diffusion time is a key parameter in order to know whether to apply the diffusion-limited or the reaction-limited model. This ratio depends on the reactivity of the pair, their respective sizes, but also on their concentrations in each phase. Despite the apparent complexity of the matter, theoretical approaches on both regimes, using simplistic approximations such as flat and two-dimensional interfaces, prove to be efficient for a large number of experimental studies using different reactive pairs. However, these theories are always treated for interfaces under quiescent conditions, which is not the case during emulsion or blend generation by turbulent mixing. Zhang \textit{et al.} \cite{Zhang2010} showed experimentally that the presence of an external flow accelerates interfacial reactions. Song \textit{et al.} \cite{Song2013} also showed an acceleration of the reaction under a compressive flow, and state that this explains the remarkable ability of co-extrusive processes to build multilayer products with little residence times. These more realistic experimental conditions are still missing in the theories of kinetic evolution of copolymer coverage of the interface.

\section{Consequences of the reactive stabilization}
\label{sec:consequences}

Reactive compatibilization of liquid dispersions is mostly used industrially because of its ability to stabilize efficiently liquid dispersions and because of the cost reduction it implies compared to the stabilization route with pre-made surfactants, since the production of tailor-made copolymers for a particular dispersion can be very expensive. However it is also worth taking an interest in the impact of the reactive stabilization route on several parameters such as the stability and the morphology of the dispersion, and the shape of the interface through the apparition of instabilities. In this section, we review these different repercussions. 

\subsection{Dispersion morphology: influence on the drop sizes}

There are several ways to mix the phases in an emulsion or a polymer blend. In order to be selective on the drop size and the role of each phase (either dispersed or continuous), one can generate the drops using microfluidic techniques or simple dripping of the dispersed phase in the continuous phase. For industrial processes however, these techniques do not allow a rapid and quantitative production, and it is then preferred to use turbulent mixing and twin-screw extrusion. These methods are based on the ability of one of the phases to deform under shear until breakup occurs to form droplets. The final drop size distribution depends in this case on both external parameters such as the temperature, the shear rate and the duration of mixing or extrusion time (though an equilibrium is rapidly achieved \cite{Sundararaj1995}), and on the fluids properties like the rheological properties of the two fluids, the interfacial tension between the two phases, and the compatibilizing agent \cite{Lee1999}. In this section, we focus only on the influence of the compatibilization process on the final drop size distribution of the polymer blend or emulsion, which is also influenced by the interfacial tension.

\begin{figure}
\begin{center}
\includegraphics[scale=0.28]{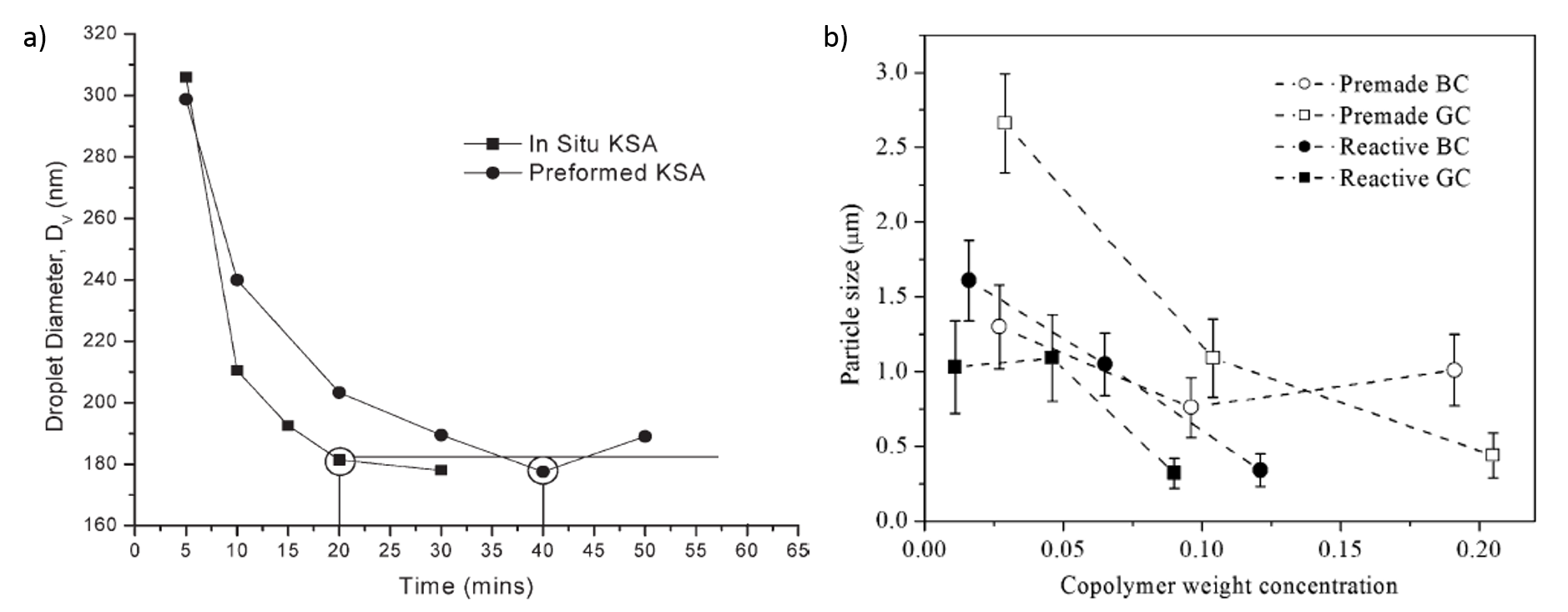}
\caption{a) Droplet size evolutions for methyl methacrylate (non organic phase) droplets in an aqueous solution of butyl acrylate prepared with a rotor/stator (3 000 rpm) using equal amounts (9.0 mmol) of preformed potassium stearate (KSA) or in situ KSA ( $\bigcirc$ minimum droplet size) (adapted from \\protectcite{ElJaby2010} with permissions). b) Particle size changes with copolymer concentration for all the compatibilized PMMA/PS blends. The terms BC and GC refer to respectively block and graft copolymers. The error bars are one standard deviations (from \protect\cite{Jeon2005} with permissions).}
\label{4.3}
\end{center}
\end{figure}

During mixing, breakup of the dispersed phase occurs to form drops, which subsequently deform to allow the formation of smaller and smaller drops until an equilibrium size is achieved. This equilibrium size is the result of the breakup and coalescence rates becoming equals. The deformation of the drop towards breakup is subject to the tangential stress tending to elongate the drop being greater than the interfacial capillary pressure \cite{Taylor1934,Wu1987} which tends to minimize the surface of the drop. This means that the lower the surface tension between the phases, the easier it is to deform the interface and we could expect smaller drop sizes. In that regard, the use of compatibilizers (preformed or in-situ generated) is relevant in order to decrease the mean drop size of the emulsion or polymer blend. But as suggested by Milner \textit{et al.} \cite{Milner1996}, the sole influence of the diminution of surface tension on the size distribution of the drops cannot account for the role of the addition of a compatibilizer, but prevention of coalescence by steric effects could. Indeed, they theoretically showed that even a low content of compatibilizer is enough to prevent coalescence of submicron-sized droplets while having only a negligible impact on the surface tension. This dual effect is shown by several studies for various systems \cite{VanPuyvelde2001}, for example nylon and ethylene-propylene rubber \cite{Wu1987,Scott1994-2,Thomas1999}, nylon and polystyrene \cite{Tan1996}, nylon and polysulfone \cite{Weber1999}, nylon and poly(methyl metracrylate) \cite{Dedecker1998}, polystyrene and poly(methyl methacrylate) \cite{Macosko1996}, polycarbonate and  styrene–acrylonitrile copolymer \cite{Wildes1999} or oil and water \cite{ElJaby2010,Ballard2015}.

However, the question remains whether the in-situ generation of the compatibilizers is more efficient than the addition of pre-formed surfactants to reduce the mean drop size. Assuming the exact same molecular structure of the compatibilizer, the equilibrium value of interfacial tension does not change between in-situ generated and pre-formed surfactants, but the kinetics of the surface coverage by the compatibilizer does. In practice, it is difficult to control the final structure of the compatibilizer during reactive compatibilization. Only few studies involve the direct comparison of the influence of either preformed or in-situ generated compatibilizers. Jeon \textit{et al.}\cite{Jeon2005} noted that polymethyl methacrylate and polystyrene blends stabilized by pre-made surfactants showed the possibility to decrease the drop size down to $\sim1$ $\mu$m, but the reactive blending allowed to go to even lower sizes ($\sim0.3$ $\mu$m) (Figure~\ref{4.3}b). This could be attributed to the dispersion of the pre-made surfactants in the form of micelles in the bulk which reduces the number of molecules adsorbed at the interface \cite{Koning1998}. Sundarajaj \textit{et al.} \cite{SundararajMacosko1995} showed that the drop size depended on the volume fraction of dispersed phase with uncompatibilized blends and blends stabilized using diblock or triblock copolymers, but not for the reactive system. They attributed this result to the complete inhibition of coalescence with the reactive stabilization. On the contrary, El-Jaby \textit{et al.}\cite{ElJaby2010} showed that in an oil-in-water emulsion, the same final drop size was obtained for both pre-formed and in-situ generated surfactants, and did not notice any influence of the amount of generated or added surfactants between the two systems on the final drop size. They showed however that the mixing time needed to reach the minimum drop size was two times longer with the pre-formed surfactant at a given shear rate (Figure~\ref{4.3}a). On the other hand, Ballard \textit{et al.} \cite{Ballard2015} showed that, for another oil-in-water system (the oil being styrene), the drop sizes were smaller for the in-situ compatibilized system than for the system with pre-made surfactants.

\begin{figure}[ht]
\begin{center}
\includegraphics[scale=0.2]{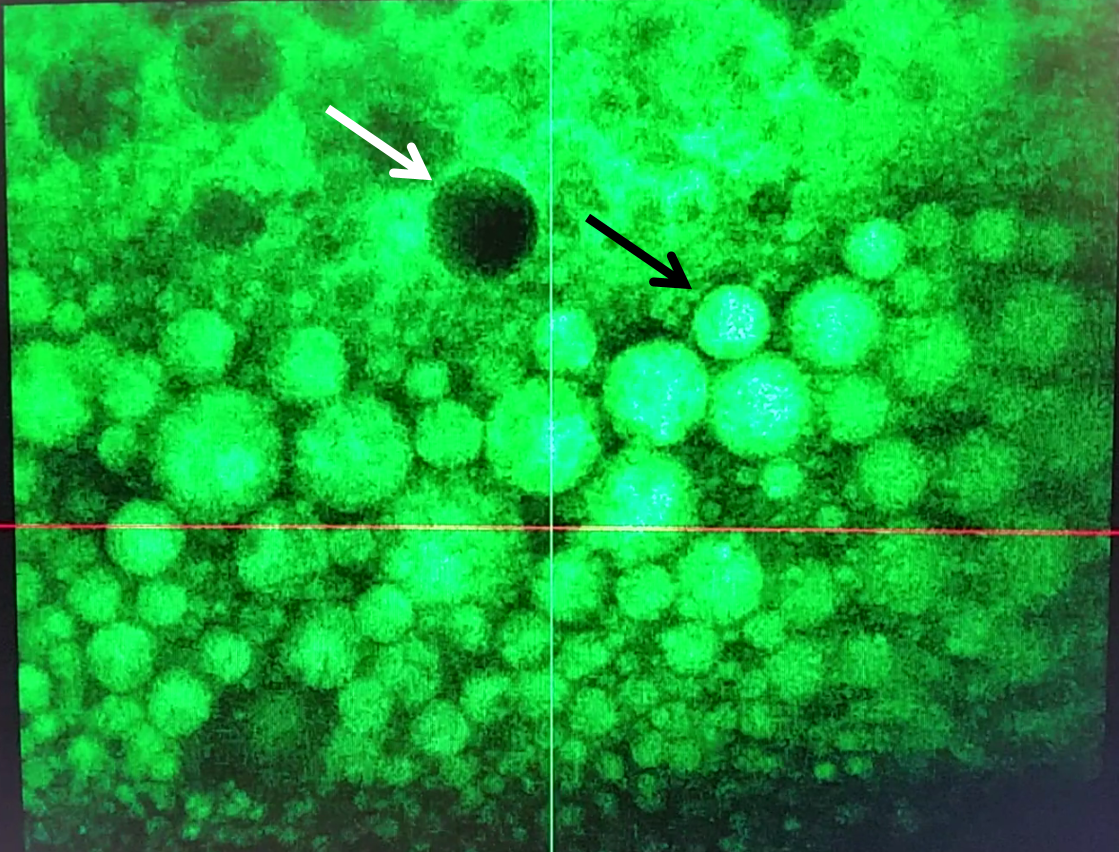}
\caption{Optical image obtained with a confocal microscope of an emulsion generated by turbulent mixing of PEG and PDMS showing both PEG-in-PDMS (black arrow) and PDMS-in-PEG (white arrow) droplets. Fluorescein was added in the PEG phase. \protect\cite{Giustiniani2016} with permissions}
\label{Mixte-emulsion}
\end{center}
\end{figure}

One drawback of the reactive stabilization is the lack of selectivity of turbulent mixing in terms of which phase will be dispersed in the other. Since no surfactants are added initially in one of the phases as it is the case for non-reactive systems, the Bancroft rule, which states that the phase in which an emulsifier is more soluble constitutes the continuous phase, does not apply. This means that if the phases also have viscosities close to each other, both phases can \textit{a priori} equally become the dispersed phase. This was observed by Giustiniani \textit{et al.} \cite{Giustiniani2016} for a PDMS/PEG emulsion where they found both PEG droplets in PDMS and PDMS droplets in PEG when preparing the emulsion by turbulent mixing (Figure \ref{Mixte-emulsion}).

\subsection{Dispersion stability}

As mentioned in~\ref{subsec:stab}, the use of pre-made surfactants, especially in the case of polymer blending, can lack ability in stabilizing the system. This is attributed to different reasons. Short molecules can diffuse easily towards the interface, but are not large enough to stay anchored at the interface \cite{Macosko1996}. Large molecules, on the contrary, are entangled enough to be anchored at the interface, but present a slower kinetic of adsorption, in particular in viscous media \cite{SundararajMacosko1995}. Moreover their ability to stabilize the blend is impaired by their lower miscibility in the bulk and their higher tendency to form micelles which reduces the number of molecules participating in the adsorption processes at the interface \cite{Koning1998}.

\begin{figure}
\begin{center}
\includegraphics[scale=0.7]{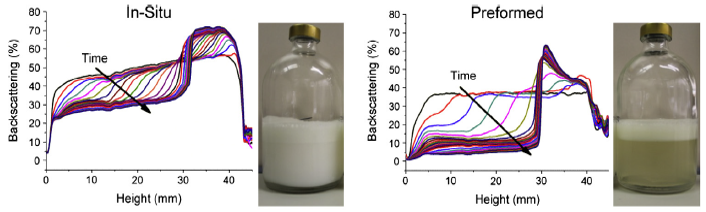}
\caption{Backscattered light intensity variation with height over time of styrene in water emulsions with identical recipes formed by either in-situ technique (left) or preformed technique (right) after leaving to cream for 1 h. The spectra are shown at 1 min intervals for the first 15 min and subsequently every 15 min until 1 h. The photographs correspond to the identical systems after allowing to cream for 1 h (from \protect\cite{Ballard2015} with permissions).}
\label{4.1}
\end{center}
\end{figure}

In order to fix these issues, reactive stabilization seems to be an appropriate candidate. Indeed, it generates the surfactant at the interface, where it is needed. This makes sure that, even though desorption of the surfactant from the interface can occur under certain conditions, there will be less micelles in the bulk, and the copolymer surface coverage at comparable amount of added pre-made surfactants and reactive molecules is higher in the reactive stabilization case \cite{Baker2001}. Ballard \textit{et al.} \cite{Ballard2015} showed that styrene-in-water emulsions were more stable using in-situ potassium oleate than preformed ones by measuring the backscattered light variation with height over time of the reactive and non-reactive emulsions (Figure~\ref{4.1}). 

Depending on the functionality of the reactive polymers in a reactive compatibilization, one can obtain several surfactant architecture, such as block copolymers, Y shaped copolymers and graft copolymers. The size of this copolymer influences its ability to stay at the interface: the larger the molecule, the less desorption from the interface occurs. By increasing the amount of reactive polymers in the bulk, Giustiniani \textit{et al.}\cite{Giustiniani2016} increased the reaction rate, and were able to form larger and larger molecules until a continuous network of cross-linked polymers was created at the interface. This allowed the complete inhibition of coalescence and Ostwald ripening between the drops and led to ultra-stable emulsions.

The reactive stabilization approach has also been used to stabilize non-polymeric emulsion. Indeed, by using surfactants (phosphatidyl cholines bearing one or two methacrylate groups) polymerizable under irradiation, Regen \textit{et al.} \cite{Regen1981} were able to stabilize vesicles and showed that these were actually more stable than liposomes which are vesicles derived from naturally occurring phospholipids. Polymerized vesicles were also generated by Fendler \textit{et al.} \cite{Fendler1984} using photo-excitation to cross-link the surfactants. Eaton \textit{et al.} \cite{Eaton1980} also reported the generation of polymerized vesicles via a chemical reaction catalysed by heat. More recently, Summers and Eastoe  \cite{Summers2003} published a review dedicated to the possible applications of polymerizable surfactants, including the formation of micelles, vesicles, bilayers and emulsions.

\subsection{Interfacial instabilities}

Instabilities at the interface have been observed experimentally under different conditions for non-reactive \cite{Shull1992,Xu1995} and reactive systems \cite{Kim2005,Kim2000,Wang2010,Jiao1999,JiaoKramer1999,Lyu1999,Zhang2005}. These emerge as strong deformations of the interface (also called fingering of the interface), sometimes leading to a mechanism called spontaneous emulsification. Kim \textit{et al.}\cite{Kim2005,Kim2000} and Jiao \textit{et al.}\cite{Jiao1999} showed that the root mean square (rms) roughening of the interface between polystyrene and polystyrene maleic anhydride increases violently for a value of copolymer coverage of the interface corresponding to the interfacial tension turning negative. Lyu \textit{et al.}\cite{Lyu1999} proposed a mechanism in which the strong decrease in interfacial tension between polystyrene and poly(methyl methacrylate), without turning negative, allows thermal fluctuations (i.e. capillary waves) to deform the interface, making it possible for other functionalized polymers to react at the interface and oversaturate it, inducing strong interfacial deformations by steric effects. Zhang \textit{et al.}\cite{Zhang2005} also observed a strong increase in rms roughening of the interface once the copolymer surface coverage exceeds its equilibrium maximum value. These enhanced roughenings of interfaces were in every case associated with an emulsification of the interface (without stirring of the system), meaning that very small droplets or even micelles detached from the interface to the bulk phase. These instabilities typically arise when the different time scales of the system are such that the population of the interface is faster than its relaxation, leading to an overpopulation of the interface and hence to what some call an effective negative interfacial tension.

\begin{figure}
\begin{center}
\includegraphics[scale=0.6]{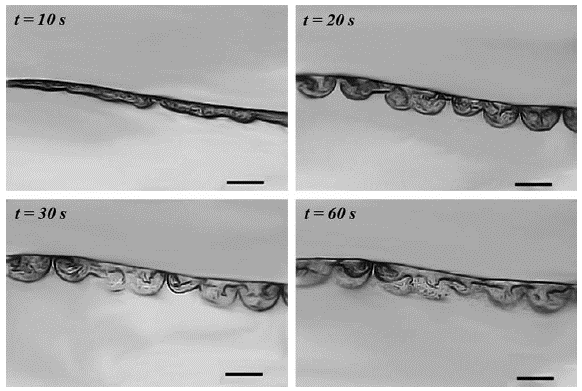}
\caption{Interfacial polymerization at 25$^{\circ}$C: top phase, 50\% BDO + 50\% PPO; bottom phase, diisocyanate. Time: a) 10 s; b) 20 s; c) 30 s; d) 60 s  (from \protect\cite{Patashinski2012}, adapted from \protect\cite{Fields1986} with permissions).}
\label{4.2}
\end{center}
\end{figure}

The formation of small droplets and micelles through interfacial instabilities induced by a negative surface tension was studied theoretically by Granek \textit{et al.}\cite{Granek1993} for a non-reactive compatibilization. More recently, the case of a reactive coupling at the interface was studied theoretically by Patashinski \textit{et al.}\cite{Patashinski2012} and via modelisation by Berezkin and Kudryavtsev \cite{Berezkin2013}. Patashinski \textit{et al.} studied the effect of a negative surface tension on a flat interface caused by overpopulation of surfactants, and successfully predicted the surface roughening observed experimentally by Fields \textit{et al.} \cite{Fields1986} (Figure~\ref{4.2}). Berezkin and Kudryavtsev studied both flat and curved interfaces and showed the influence of the size of each of the reactive species on the final morphology of the blend.

Reactive stabilization offers many advantages compared to the use of pre-made surfactants: faster kinetics of surface coverage leads to a faster reduction of interfacial tension and inhibition of coalescence, both having an impact on the final drop size under mixing. Moreover it allows to create insoluble surfactants at the interface, which could not be used in pre-made form, and can lead to super-stable systems. However, the apparition of interfacial instabilities which roughen the interface between the two phases is very often observed, but can be used as a way to produce micelles in a controlled way \cite{Bachmann1991}.

\section{Conclusion}

The stabilization of emulsions and polymer blends is made possible by the presence of surfactants (both pre-made or created in situ) at the interface. Their kinetics of adsorption/creation and desorption play an important role in the stabilization process and can be accessed by measuring the evolution of interfacial tension or surfactant surface coverage between the phases in contact. Our knowledge on these for non-reactive interfaces is now fairly advanced, thanks to theoretical models and numerous experimental studies in the literature. However, because of the difficulty for large molecules to diffuse, non-reactive stabilization often results in poor stabilization when dealing with viscous phases.  

To deal with these issues, a number of studies used a reactive stabilization, which forms the surfactant directly at the interface, where it is needed. It avoids the solubility issues of the large molecules in the bulk which prevents the addition of enough surfactants for the stabilization. Depending on the structure of the reactive molecules, it can also, under certain conditions, block the desorption of the surfactants. The size of the reactive molecules is also necessarily smaller than the pre-made molecule it replaces, meaning lower diffusion times and therefore faster kinetics of surface coverage by surfactants.

For reactive interfaces, the values of interfacial tension (or equivalently surfactant surface coverage) at long times ("equilibrium" values) seems to follow the same behaviour as for their non-reactive counterparts, although systematic comparisons between the two are to this day limited in number. Most of the studies involving reactive stabilizations take the uncompatibilized blend or emulsion as reference, and make no mention of the non-reactive stabilization process.

The kinetics of evolution of surface tension at reactive interfaces, however, has received a lot of attention these past two decades. Various theories have emerged, differentiating two regimes: reaction-controlled when the reactivity of the functionalized molecules and/or the concentration of functionalized groups are low, and diffusion-controlled when the reaction is fast and the only process limiting the kinetics of evolution is the diffusion of the reactive molecules. Numerous experimental studies report either the former or the latter, but the conclusions are sometimes a matter of debate \cite{Oyama2001}. Also, these theories are intentionally simplistic (flat, two-dimensional and quiescent interfaces, no surfactant desorption, etc.). Yet, even if they seem to be able to capture the general behaviour of the evolution of interfacial tension, reactive stabilizations are mostly used in industrial processes implying either co-extrusion or turbulent mixing, so they could be improved to match the experimental conditions. 

Another important aspect to understand would be the apparition of instabilities at reactive interfaces. Only a few studies are tackling this phenomenon, though it could have a great impact in industrial processes where roughening of interfaces is to be avoided, or inversely provide new ways of making both micro and nanoemulsions.

Finally, it seems adequate to emphasize that emulsions and polymer blends are usually treated separately, when they are physically the same class of materials, the only difference being the size of the bulk molecules. Here for example, we handled them both equally regarding interfacial tension, and surely other aspects might be treated with the same approach. We certainly expect both fields to benefit from a closer collaboration.

\section*{Acknowledgements}
The authors would like to thank Dominique Langevin, Anniina Salonen, Emmanuelle Rio, François Boulogne, Liliane Léger and Fréderic Restagno for numerious in-depth discussions on this subject. We acknowledge funding from the European Research Council (ERC) under the European Union’s Seventh Framework Program (FP7/2007-2013) in form of an ERC Starting Grant, agreement 307280-POMCAPS.

%\bibliography{Biblio} 
%\bibliographystyle{unsrt}

\end{document}